# Radiative conductivity and abundance of post-perovskite in the lowermost mantle

Sergey S. Lobanov[1,2]*†, Nicholas Holtgrewe[1,3],†, Jung-Fu Lin[4,5], Alexander F. Goncharov[1,6]

[1]Geophysical Laboratory, Carnegie Institution of Washington, Washington, DC 20015, USA

[2]V.S. Sobolev Institute of Geology and Mineralogy SB RAS, Novosibirsk 630090, Russia

[3]Howard University, 2400 Sixth Street NW, Washington, DC 20059, USA

[4]Department of Geological Sciences, Jackson School of Geosciences, The University of Texas at Austin, Austin, TX 78712, USA

[5]Center for High Pressure Science and Technology Advanced Research (HPSTAR), Shanghai 130012, China

[6]Key Laboratory of Materials Physics, Institute of Solid State Physics CAS, Hefei 230031, China

*Corresponding author: slobanov@carnegiescience.edu

†These authors contributed equally to this work

## Abstract

Thermal conductivity of the lowermost mantle governs the heat flow out of the core energizing planetary-scale geological processes. Yet, there are no direct experimental measurements of thermal conductivity at relevant pressure-temperature conditions of Earth's core-mantle boundary. Here we determine the radiative conductivity of post-perovskite at near core-mantle boundary conditions by optical absorption measurements in a laser-heated diamond anvil cell. Our results show that the radiative conductivity of $Mg_{0.9}Fe_{0.1}SiO_3$ post-perovskite (< 1.2 W/m/K) is ~ 40% smaller than bridgmanite at the base of the mantle. By combining this result with the present-day core-mantle heat flow and available estimations on the lattice thermal conductivity we conclude that post-perovskite is as abundant as bridgmanite in the lowermost mantle which has profound implications for the dynamics of the deep Earth.

## Main text

The lowermost 200-400 km of the mantle is a critical region responsible for the core-mantle interaction powering all major geological processes on Earth[1]. Specifically, thermal conductivity of the thermal boundary layer (TBL) above the core-mantle boundary (CMB) determines the heat flow out of the core that provides energy to sustain the mantle global circulation and to drive the geodynamo[1, 2]. Seismic structures of the lowermost mantle, however, are complex[3, 4] implying that the thermal conductivity of the region is non-uniform due to variations in chemical (*e.g.* iron) and/or mineralogical contents as well as texturing of the constituting minerals. The nature of the seismic heterogeneity near the CMB, including a sharp increase in shear wave velocity and anti-correlations of seismic parameters, has been linked to the bridgmanite (Bdgm) to post-perovskite (Ppv) transition[5, 6], as these phases have contrasting elastic, rheological, and transport properties (*e.g.* Ref.[6]). Indeed, measurements and computations of lattice thermal conductivity ($k_{lat}$) in Bdgm and Ppv revealed that Ppv conducts heat 50-60 % more efficiently than Bdgm[7, 8], suggesting that the distribution of the Ppv phase can significantly enhance the heat flux out of the core. However, no mineral physics constraints are available on the radiative thermal conductivity ($k_{rad}$) of Ppv, which should play an increasingly important role at high temperature[9], as well as on the Ppv abundance in the lowermost mantle. This has hampered our understanding of how the heat flux across the CMB may vary laterally and what magnitude of the energy source in the mantle and the outer core is needed to power their convections.

Previous estimates of radiative thermal conductivity at lower mantle conditions were based on high-pressure room-temperature measurements of the absorption coefficients of representative minerals in the mid/near-infrared and visible spectral range[10-14]. The presence of an intense thermal radiation emitted from the hot sample makes measurements of the optical properties at high temperatures relevant to the lowermost mantle (approximately 3000 K) very challenging as common light sources have similar radiative temperatures. In the absence of the experimental data, the effect of temperature was neglected. It has been argued that temperature-induced variations in the absorption spectrum of Bdgm are small as intensity of the crystal-field band is determined by a symmetry distortion of the $Fe^{2+}O_{12}$-polyhedra in Bdgm[12]. However, the intensity of the crystal field spectrum in iron-bearing minerals can be sensitive to pressure, temperature, iron concentration, and iron spin states[10, 15-19]. Moreover, due to a substantial amount of $Fe^{3+}$ in Bdgm and Ppv at lower mantle conditions[20, 21], the absorption coefficients and thus the radiative conductivity of these minerals are also governed by the $Fe^{2+}$-$Fe^{3+}$ charge transfer (CT)[15], which is expected to diminish with temperature[22]. Therefore, to ascertain the presently unknown radiative conductivity of iron-bearing minerals at expected CMB conditions, it is of fundamental importance to underpin physical mechanisms that govern optical absorption of lower mantle minerals at simultaneous high pressure-temperature conditions. It is also desirable to assess the intensity of the absorption bands as a function of the total iron content because the solubility of iron in lower mantle minerals can vary with pressure, temperature, and phase/spin transitions[23].

In this work we have studied optical properties of iron-rich, $Fe^{3+}$-bearing Ppv ($Mg_{0.6}Fe_{0.4}SiO_3$) at 300-2050 K and 130 GPa using laser-heated diamond anvil cells combined with a pulsed ultra-bright supercontinuum probe (400-2400 nm) synchronized with the collection windows of gated visible and infrared detectors. The latter allowed diminishing the contribution of thermal radiation to the optical signal at high temperature by several orders of magnitude. We determined quantitatively the reduction in intensity of all Ppv absorption bands in the visible-infrared range at high pressure and temperature. These, in turn, have allowed us to constrain the spin and valence states of iron in Ppv and its radiative conductivity at expected CMB conditions. Combined with literature data on the absorption coefficients of Bdgm and ferropericlase (Fp), these results enable us to construct a radiative thermal conductivity model for the lowermost mantle in order to address the CMB heat flux and Ppv content in the region.

Iron-rich ($Mg_{0.6}Fe_{0.4}SiO_3$) Ppv was synthesized in a diamond anvil cell using enstatite powder of the corresponding composition sandwiched between NaCl layers, similar to procedures used in previous reports on the spin and valence states in Ppv[24]. The sample was compressed to 130 GPa at room temperature and then laser heated at 2500-2800 K for 6-8 hours at 13IDD beamline of GSECARS, Advanced Photon Source. The synthesis was confirmed by synchrotron x-ray diffraction that showed a complete transformation to the Ppv phase. After successful synthesis of PPv, analysis of the measured room-temperature absorption spectrum in the 6000-22500 $cm^{-1}$ range shows three distinct features at energies typical for crystal field and CT transitions (Fig. 1)[15].

The band assignment of our absorption spectra is based on first-principles calculations at zero Kelvin[25, 26] and is consistent with experimental room-temperature Mőssbauer data collected previously on samples with similar synthesis conditions[24], providing necessary reference information on the identification for the iron spin and valence states in Ppv at 300 K (see Supplementary Materials). Specifically, low spin (LS) $Fe^{3+}$ occupies the 6-fold B-site (Supplementary Fig. S1), while the 8-fold A-site is occupied by both high spin (HS) $Fe^{3+}$ and HS $Fe^{2+}$. Combining these results with Tanabe-Sugano diagrams for $d^5$ and $d^6$ elements we deduce the spin-allowed electronic transitions in the crystal field of iron in Ppv at 130 GPa and 300 K: $^5E_g \rightarrow {}^5T_{2g}$ in A-site HS $Fe^{2+}$ and $^2T_{2g} \rightarrow {}^2T_{1g}$ ($^2A_{2g}$) in B-site LS $Fe^{3+}$. Importantly, all crystal field transitions in HS $Fe^{3+}$ are spin-forbidden.

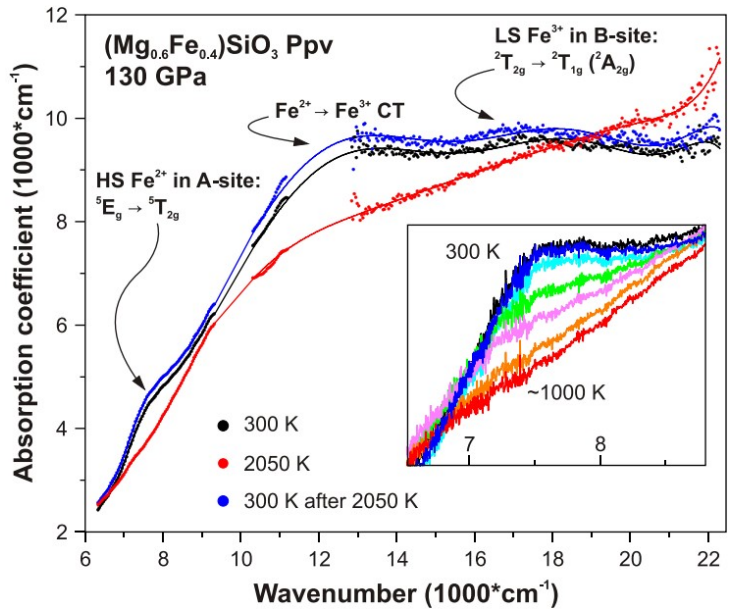

**Figure 1.** Absorption coefficients of Ppv with $Mg_{0.6}Fe_{0.4}SiO_3$ composition at 130 GPa. Black: before heating; red: at ~ 2050 K; blue: quenched to room temperature. Spectral regions that lack absorbance data are due to the need to block the spectral range (9500-10500 cm$^{-1}$) near the heating laser wavelength (1070 nm) and the limited sensitivity of the iCCD and InGaAs detectors at 11000-13000 cm$^{-1}$. Lines are fits to the spectra to guide the eye. Inset: temperature-induced changes in the 6500-8700 cm$^{-1}$ range at T = 300 (black and blue) to ~ 1000 K (red). All observed transformations are reversible.

The band centered at ~ 7500 cm$^{-1}$ is assigned to the $^5E_g \rightarrow {}^5T_{2g}$ transition in HS $Fe^{2+}$ at the A-site based on the relatively weak crystal field splitting energy of its 8-fold coordination (see Supplementary Materials). Upon laser-heating, the $^5E_g \rightarrow {}^5T_{2g}$ band red-shifts and weakens continuously with increasing temperature to T ~ 1000 K (Fig. 1). Interestingly, the red-shift represents a typical behavior of a crystal field band due to the thermal expansion decreasing the crystal field splitting energy. The vanishing intensity, however, is highly uncommon for single-ion crystal field bands[15]. Likewise, the band intensity is not scaled linearly with the Ppv iron content (Supplementary Fig. S2), as would be expected for single-ion crystal field bands, suggesting that the absorption of light by the A-site HS $Fe^{2+}$ is enhanced by interactions with the edge-sharing B-site $Fe^{3+}$. Indeed, exchange-coupling interactions of adjacent transition metal ions have been employed to explain the anomalous temperature and concentration behavior of the crystal field bands[27, 28]. However, the importance of this transition mechanism for the absorption coefficient of lower mantle minerals has been previously overlooked. Alternatively, the disappearance of the band at T ~ 1000 K may indicate that the $^5E_g \rightarrow {}^5T_{2g}$ band becomes parity-forbidden due to temperature-induced static and/or dynamic distortions imposing a center of symmetry to the A-site (vibronic decoupling)[15]. A detailed crystallographic analysis of intensities and positions of multigrain Bragg reflections may allow discriminating between these two hypotheses. In the absence of such information we assume the exchange-coupled mechanism plays a dominant role in the intensity weakening.

The band centered at 17500 cm$^{-1}$ is assigned to the $^2T_{2g} \rightarrow {}^2T_{1g}\,({}^2A_{2g})$ transition in LS Fe$^{3+}$ at the B-site as it corresponds to the lowest energy spin-allowed excited state of $d^5$ elements (Supplementary Fig. S3 and Supplementary Materials). At 2050 K, the LS Fe$^{3+}$ band is no longer observed (Fig. 1), indicating a temperature-induced LS to HS transition at expected P-T conditions of the CMB. In Al-bearing Bdgm all iron is likely accommodated by the 12-folded A-site and remain in the HS state at lower mantle P-T (Refs.[29]), as opposed to Fp which may contain a significant portion of LS iron (up to 50 %) at the bottom of the lower mantle[30]. Accordingly, Fp is the only major host of LS iron in the lower mantle.

The most prominent spectral feature observed here is the broad absorption band at 12000-13000 cm$^{-1}$ (Fig. 1) which we assign to a Fe$^{2+}$-Fe$^{3+}$ CT transition. This band is characteristic of mixed-valence compounds[15] and has an increased bandwidth compared to crystal field bands[15]. The decreased intensity at high temperature supports this assignment because the number of absorbing Fe$^{2+}$-Fe$^{3+}$ pairs in the ground state ($N$) decreases with temperature ($T$) as $N \sim 1-\exp(-E_a/kT)$, where $E_a$ is the activation energy for a thermally-induced CT and $k$ is the Boltzmann constant. Additionally, Fe$^{2+}$-O and Fe$^{3+}$-O CT (absorption edge) may give rise to the intense absorption in the spectrum at frequencies above 22000 cm$^{-1}$ (Fig. 1; Supplementary Fig. S2).

A conventional approach to assess $k_{rad}$ for a given room temperature absorption coefficient (α) is:

$$k_{rad}(T) = \frac{4n^2}{3} \int_0^\infty \frac{1}{\alpha(v)} \frac{\partial I(v,T)}{\partial T} dv \quad (1)$$

where *n* is the refractive index of a given mineral, *v* is frequency, and *I(v, T)* is the Planck function[11, 12]. As revealed in this study, the absorption coefficient of Ppv is controlled by a number of physical mechanisms such as the temperature-activated spectroscopic selection rules and temperature-inhibited magnetic coupling. It follows that most accurate values of $k_{rad}$ require absorption coefficients evaluated at high temperature. Unlike previous studies of lower mantle $k_{rad}$ (Refs.[10-14]), we determined the efficiency of the radiative heat transfer mechanism at high temperature based on the 2050 K absorption data. The comparison based on the 300 and 2050 K data (Supplementary Fig. S4) reveals that the effect of temperature-dependent absorption coefficient on $k_{rad}$ is on the order of 10 %, in spite of the substantial changes in the absorption spectra (Fig. 1). The relative difference tends to get smaller with increasing temperature due to the diminishing overlap of the blackbody radiation with Ppv absorption. As the refractive index of Ppv at 130 GPa is unknown, we assume a wavelength-independent value of *n* = 2.1, based on a density-corrected refractive index of Bdgm[31]. We estimated a combined uncertainty of ~ 25 % in $k_{rad}$ due to uncertainties in determination of both sample thickness and refractive index (see methods).

The radiative conductivity of the lowermost mantle is determined by absorption coefficients of its constituting materials likely including Ppv, Bdgm, and Fp. To construct a model of radiative conductivity in the TBL we used previously reported room-temperature absorption coefficients of Ppv $(Mg_{0.9}Fe_{0.1}SiO_3)$[19], Bdgm $(Mg_{0.9}Fe_{0.1}SiO_3)$[13], and Fp $(Mg_{0.85}Fe_{0.15}O)$[13]. As revealed in this study (Supplementary Fig. S4), the use of 300 K Ppv absorption coefficients underestimates $k_{rad}$ by ~ 10 % at CMB conditions (P ~ 135 GPa and T ~ 4000 K). To address the effect of iron substitution on radiative conductivity, we applied the 10 % correction determined in this work to $k_{rad}$ values inferred from the 300 K absorption data of Ppv with variable iron content (10-30 %) reported by Goncharov et al.[19] (Fig. 2 inset), assuming that iron-depleted Ppv exhibits a similar temperature-dependence of the absorption coefficient. Given the similarities in iron valence and spin distribution between the A- and B-sites in the structures of Bdgm and Ppv (Refs.[25, 26]), one may expect that the light absorption mechanisms in Bdgm have an analogous temperature-dependence to that in Ppv. Accordingly, we applied the same 10 % upward correction to Bdgm radiative conductivity.

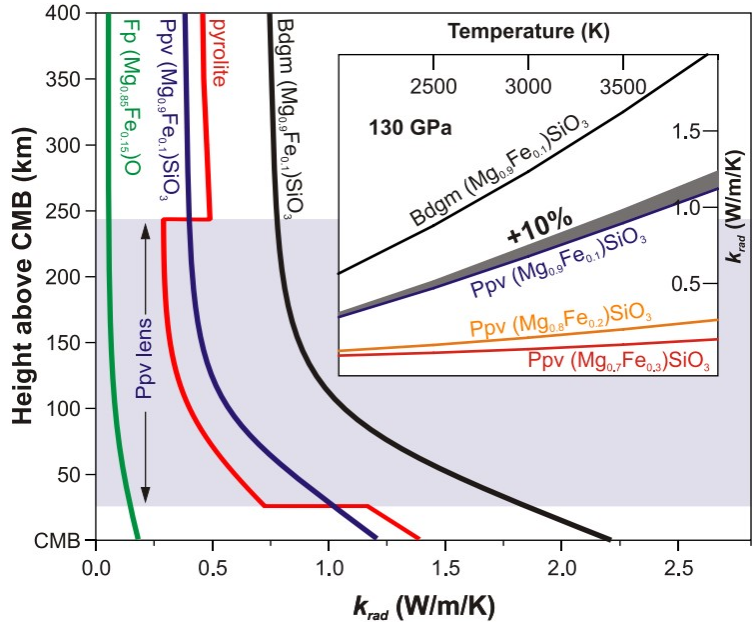

**Figure 2.** Radiative thermal conductivity ($k_{rad}$) of Ppv (dark blue) and Bdgm (black) with $Mg_{0.9}Fe_{0.1}SiO_3$ composition in the thermal boundary layer (TBL) along the intermediate geotherm of Ref.[4]. $k_{rad}$ of Ppv was calculated using its absorption coefficient at 130 GPa reported by Goncharov et al.[19] within the conventional formula (*e.g.* Ref.[11]) and then corrected ($k_{rad}$*1.1) for high temperature effects on the absorption coefficient. Pressure-dependence was neglected as it is weak over the relatively small pressure range of TBL. The overall uncertainty in $k_{rad}$ is ~ 25 % due to the ambiguity in the refractive index and sample thickness of Ppv. $k_{rad}$ of Bdgm is based on the spectrum at 133 GPa and 300 K reported by Goncharov et al.[11] and was corrected for high-temperature effects ($k_{rad}$*1.1). $k_{rad}$ of Fp, $(Mg_{0.85}Fe_{0.15})O$, (green) is calculated based on the room-temperature spectrum at 133 GPa (Ref.[13]) and is not corrected for temperature effects. The red curve is the Hashin-Shtrikman averaged[32] $k_{rad}$ of a pyrolite composition (0.8 Bdgm or Ppv, and 0.2 Fp by volume). The kinked form of the red curve is due to the 'double-crossing'[33] in the Ppv lens (gray area) assuming a temperature gradient of 7 K/km and a

CMB temperature of 4000 K. Inset: radiative conductivity of $Mg_{1-x}Fe_xSiO_3$ - Ppv for x = 0.1 (dark blue), x = 0.2 (orange), and x = 0.3 (red), respectively at 130 GPa. Blue-colored area shows the 10 % correction due to the temperature-dependent absorption coefficient.

Figure 2 shows a model of the radiative thermal conductivity in the lowermost mantle assuming a ~7 K/km temperature gradient, CMB temperature of 4000 K, and 10 % iron in Bdgm and Ppv (Ref.[34]). According to the model, the ability of Ppv to conduct heat by radiation is ~ 40 % lower than that of Bdgm with a similar composition at expected CMB conditions of 135 GPa and 4000 K. The radiative conductivity of Ppv increases from ~ 0.35 at ~ 400 km above CMB to 1.2 W/m/K at CMB. Over the same depth interval, Bdgm $k_{rad}$ increases from 0.75 to 2.2 W/m/K, while in Fp $k_{rad}$ is < 0.2 W/m/K, a much lower value than that of Ppv and Bdgm. Radiative conductivity of Ppv, Bdgm, and Fp at CMB conditions is lower than the lattice thermal conductivity of their iron-free counterparts: 16.8 W/m/K (Ppv), 9 W/m/K (Bdgm), ~ 20 W/m/K (periclase) at corresponding conditions, according to experimental measurements using the thermoreflectance method[7, 35]. However, with increasing iron contents in these lower-mantle minerals, the lattice thermal conductivity is expected to decrease significantly due to the well-known phonon scattering effects in their lattices[8, 35-37], enhancing the role of the radiative thermal conductivity near CMB.

Combining the radiative conductivity model (Fig. 2) with the previously reported $k_{lat}$ values of iron-free Bdgm and iron-free Ppv, we have constrained the total thermal conductivity of a pyrolytic mantle at near CMB conditions to 12.2 and 18.5 W/m/K for Bdgm- and Ppv-dominated rock, respectively. These values, however, are upper bounds on the thermal conductivity because the substitution of iron and/or aluminum in Ppv and Bdgm would inevitably decrease their $k_{lat}$ by 10-50 % due to the phonon scattering effects (Refs.[8, 13, 14, 35-37]). Accordingly, the total thermal conductivity is

$$k = \omega k_{lat}^{Fe-fr\ pyrolite} + k_{rad}^{pyrolite} \quad (2)$$

where $\omega$ is the downward correction factor for $k_{lat}$ in the pyrolite model due to the impurity-induced phonon scattering. The lattice and radiative components for the pyrolytic mantle can thus be modeled using the following equations:

$$k_{lat}^{pyrolite} = \omega z \langle 0.8 k_{lat}^{Fe-free\,Ppv} + 0.2 k_{lat}^{periclase} \rangle + \omega(1-z)\langle 0.8 k_{lat}^{Fe-free\,Bdgm} + 0.2 k_{lat}^{periclase} \rangle \ (3)$$

$$k_{rad}^{pyrolite} = z\langle 0.8 k_{rad}^{Ppv} + 0.2 k_{lat}^{Fp} \rangle + (1-z)\langle 0.8 k_{lat}^{Bdgm} + 0.2 k_{lat}^{Fp} \rangle \ (4)$$

where $z$ is the abundance of Ppv, $z = V_{Ppv}/(V_{Ppv} + V_{Bdgm})$, where $V_{Ppv}$ and $V_{Bdgm}$ are the volumetric fractions of Ppv and Bdgm, respectively, and $\langle \ \rangle$ brackets denote the Hashin-Shtrikman averaging used to derive the effective conductivity of the mixture[32]. The fraction of Fp in the lower mantle is typically assumed to be on the order of 0.2 (Ref.[38]) for a pyrolite mantle, but the global abundance and lateral variations in Ppv are largely unconstrained. Seismic studies have revealed patches at 200-400 km above the CMB with an S-wave velocity increase of a few percent that may be indicative of the presence of Ppv (Ref.[39]). At the same time, it has been proposed that the pressure range of the Bdgm-Ppv transition is inconsistent with the depth of the D'' seismic discontinuity, questioning the presence of Ppv in the lowermost mantle[40, 41]. It is thus of great interest to the deep-Earth community to constrain the abundance of Ppv in the TBL in order to evaluate the geodynamic consequences of the Bdgm-Ppv phase transition and the associated increase in the heat flow out of the core ($Q_{CMB}$)[3, 4, 33].

Geodynamic modelling of the mantle and core has provided an estimate of $Q_{CMB} = 13 \pm 3$ TW (Ref.[2, 39, 42]), which is related to the temperature gradient and thermal conductivity of the TBL by the Fourier law of heat conduction:

$$Q_{CMB} = A_{CMB} k \Delta T \ (5)$$

where $A_{CMB}$ is the surface area of the CMB and $\Delta T$ is the temperature gradient above the CMB. Substituting Eqn. (2) for $k$ in Eqn. (5) and assuming an average temperature gradient of 7 K/km (Ref.[34]) we can model the amount of Ppv in a pyrolytic mantle required to sustain the present-day $Q_{CMB}$ as a function of the downward correction factor ($\omega$) to the lattice conductivity (Fig. 3). The values of the thermal conductivity for Ppv, Bdgm, and Fp used in our model are given in the Supplementary Table S1. Seismic data suggests that the distribution of Ppv in the lowermost mantle is highly non-uniform, with Ppv concentrated in large (up to 2000 km wide and ~ 200 km thick) lens-shaped structures[3, 4, 39]. Accordingly, we applied the Hashin-Shtrikman averaging[32] in Eqns. (3, 4) in order to model the thermal conductivity of pyrolytic mantle in the TBL layer (Fig. 3). The Hashin-Shtrikman approach proved adequate in determining the effective thermal conductivity of two-phase composites in the absence of detailed information on the spatial distribution of the phases[32].

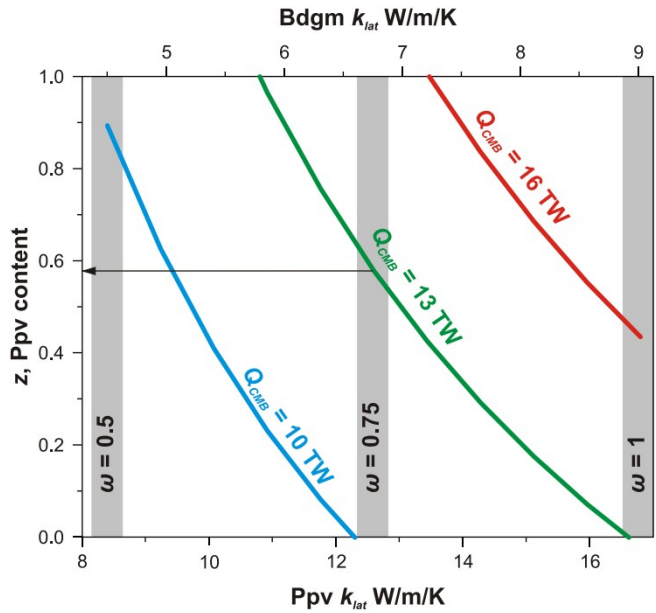

**Figure 3.** Ppv content in the thermal boundary layer (TBL) required to sustain the CMB heat flow of 10, 13, and 16 TW (blue, green, and red, respectively) for a given downward correction factor in the lattice conductivity ($\omega \leq 1$) at a temperature gradient of 7 K/km (see text for details). Ppv content ($z$) is defined as $z = [V_{Ppv}/(V_{Ppv}+V_{Bdgm})]$, where $V_{Ppv}$ and $V_{Bdgm}$ are the volumetric fractions of Ppv and Bdgm, respectively. Black arrow show the estimation of the Ppv content in the TBL assuming that the correction factor is at 25 % according to molecular dynamic simulations[8, 37] and $Q_{CMB}$ = 13 TW.

Our model suggests that Ppv should be present in the lowermost mantle to sustain 13 TW of the present-day CMB heat flow. For instance, a modest downward correction in the lattice thermal conductivity ($\omega = 0.75$) for the pyrolite model requires ~ 1:1 ratio of Ppv and Bdgm (Fig. 3). The inferred abundance of Ppv is based on a model that does not rely on the highly uncertain parameters such as the depth and Clapeyron slope of the Bdgm-Ppv transition. Nonetheless, the derived volume ratio of Bdgm/Ppv of approximately 1 is consistent with the value derived from the analysis of the D'' seismic discontinuity in the region[43]. The presence of such large portions of Ppv with relatively low thermal conductivity at the base of the lower mantle would result in mantle temperatures up to 500 K higher than that in a Ppv-free lowermost mantle, decreasing the buoyancy of plumes[44]. Likewise, higher temperatures will reduce mantle viscosity amplifying the vigor of whole mantle convection and triggering small-scale convection in the TBL[45]. If the lowermost mantle is enriched in iron, the downward correction in $k_{lat}$ may be as large as 0.5 (Refs.[35, 36]), in which case the TBL must be dominated by Ppv even for the $Q_{CMB}$ of 10 TW. On the other hand, $Q_{CMB} > 16$ TW demand that the whole TBL region is dominated by Ppv even for $\omega = 1$. However, the $Q_{CMB} < 10$ TW and $Q_{CMB} > 16$ TW scenarios are unlikely because they contradict mineral physics data on the impurity-dependence of lattice thermal conductivity of lower mantle minerals[8, 35-37]. Overall, our models show that most recent estimates of $Q_{CMB} = 13 \pm 3$ TW (Ref.[2, 39, 42]) are consistent with the Ppv interpretation of the D'' seismic discontinuity. Future experimental and theoretical studies on the impurities effects on $k_{lat}$ will enhance our understanding of the mineralogy and dynamical consequences of the Bdgm-Ppv phase transition of the CMB region.

## Acknowledgements

This work was supported by the NSF Major Research Instrumentation program DMR-1039807, NSF EAR-1015239, NSF EAR-1520648 and NSF EAR/IF-1128867, the Army Research Office (56122-CH-H), the National Natural Science Foundation of China (grant number 21473211), the Chinese Academy of Sciences (grant number YZ201524), the Carnegie Institution of Washington and Deep Carbon Observatory. A. F. G. was partly supported by Chinese Academy of Sciences visiting professorship for senior international scientists (Grant No.

2011T2J20 and Recruitment Program of Foreign Experts. S.S.L. was partly supported by the Ministry of Education and Science of Russian Federation (No. 14.B25.31.0032). We acknowledge J. Liu, J. Yang, and V. Prakapenka for helping with the preparation and synthesis of the Ppv samples. The sample synthesis and XRD experiments were conducted at GSECARS 13IDD beamline of the Advanced Photon Source, Argonne National Laboratory. GeoSoilEnviroCARS (Sector 13), Advanced Photon Source (APS), Argonne National Laboratory is supported by the National Science Foundation – Earth Sciences (EAR-1128799) and Department of Energy – GeoSciences (DE-FG02-94ER14466). This research used resources of the Advanced Photon Source, a U.S. Department of Energy (DOE) Office of Science User Facility operated for the DOE Office of Science by Argonne National Laboratory under Contract No. DE-AC02-06CH11357. J. F. Lin acknowledges support from the US National Science Foundation (EAR-1446946).

# SUPPLEMENTARY MATERIALS AND METHODS

## Methods

**Absorption measurements** in the visible (13000-22000 cm$^{-1}$) and infrared ranges (6200-11000 cm$^{-1}$) were divided into heating runs at two laser-heating systems equipped with different spectrometers and detectors (denoted hereafter as VIS and IR runs). A Leukos Pegasus pulsed supercontinuum light source (400-2400 nm) was inserted into the optical path of both systems, serving as a probe. For the VIS, the transmitted probe light was collected using a spectrometer with a 300 gr/mm grating and 300 nm focal length and a gated iCCD detector (Andor iStar SR-303i-A) synchronized with the supercontinuum pulses, identically to that described in Ref.[1]. For the IR, we used Action Spectra Pro 2300i spectrometer equipped with a 150 gr/mm grating and an ungated InGaAs detector (Princeton Instruments Model 7498-0001). The collection of 2500 supercontinuum pulses over 10 ms was initiated 200 ms after the start of the 1 s laser heating. The large number of bright probe pulses in a short accumulation window was important to increase signal-to-noise ratio and suppress thermal background. The gap in absorption coefficients at ~9000-10500 cm$^{-1}$ is due to the filters used to block the 1070 nm heating pulse, while the gap at 11000-13000 cm$^{-1}$ is due to the limited sensitivity of the iCCD and InGaAs detectors. The sample absorbance in these ranges was extrapolated based on the room temperature data collected with a conventional light source.

**Double-sided laser-heating and temperature measurements** in the VIS runs were identical to that of Ref.[1]. For the IR runs, at double-sided laser-heating we could only perform single-sided temperature measurements for which the grating was centered at 1320 nm and the supercontinuum was blocked. In both VIS and IR, temperatures measurements were taken before and after the corresponding collection of the absorbance data at an identical laser power. The laser power was optimized to achieve comparable temperatures in the VIS and IR runs (within 50-100 K). Optical responses of the VIS and IR systems were calibrated using a standard white lamp (Optronics Laboratories OL 220C). To extract the temperature, the emission spectra were fitted to the Planck black body function using the T-Rax software (C. Prescher). The reported temperature is an average of the VIS and IR measurements.

**Sample thickness** required to deduce the absorption coefficient of $Mg_{0.6}Fe_{0.4}SiO_3$ - Ppv was determined via white light interferometry at 130 GPa through the NaCl pressure medium. Additionally, the sample thickness was confirmed by direct SEM imaging of the recovered sample. The difference in sample thickness of ~ 20 % between the two estimates is the main source of uncertainty in $k_{rad}$ (~ 25 %). Other contributions to the uncertainty include ambiguities in the refractive index of Ppv.

## SUPPLEMENTARY TEXT

**Valence and spin state of iron in $Mg_{0.6}Fe_{0.4}SiO_3$ post-perovskite (Ppv)**

The interpretation of synchrotron-Mőssbauer (SMS) and X-ray emission spectroscopy (XES) data on Ppv samples synthesized from polycrystalline enstateite with identical or closely-related composition[2-4] is non-unique. However, density functional theory + Hubbard U allowed computing iron quadrupole splittings for Ppv[5] that, in turn, allowed a consistent interpretation of the experimental data. Based on the combined theory-experiment view of the valence and spin state of iron in Ppv, we presume that iron in our $Mg_{0.6}Fe_{0.4}SiO_3$ Ppv is distributed between the A- (~ 70 % high-spin $Fe^{2+}$ and ~ 10 high-spin $Fe^{3+}$) and B-sites (~ 20 % low-spin $Fe^{3+}$).

**Band assignment in the absorption spectrum of $Mg_{0.6}Fe_{0.4}SiO_3$ Ppv**

Our band assignment is consistent with the theory-experiment view in the presencse of high spin $Fe^{2+}$ in a weak crystal field (large A-site), the presence of both $Fe^{2+}$ and $Fe^{3+}$ in the synthesized sample, and low spin $Fe^{3+}$ in a strong crystal field (small B-site).

Specifically, $^5E_g \rightarrow {}^5T_{2g}$ in $Fe^{2+}$ transitions at 6-folded sites (stronger crystal field) typically show energies of 10000-12000 cm$^{-1}$, while in bridgmanite a 12-folded (weaker crystal field) HS $Fe^{2+}$ has a $^5E_g \rightarrow {}^5T_{2g}$ transition at ~ 7000 cm$^{-1}$ (Ref.[6]). Accordingly, we assign the band centered at ~ 7500 cm$^{-1}$ to the $^5E_g \rightarrow {}^5T_{2g}$ in $Fe^{2+}$ at the A-site.

Tanabe-Sugano diagram for $Fe^{3+}$ at the B-site (Supplementary Fig. S3) is consistent with the assignment of the band at ~ 17500 cm$^{-1}$ to the $^2T_{2g} \rightarrow {}^2T_{1g} ({}^2A_{2g})$ transition in LS $Fe^{3+}$. The assignment can be tested by supposing the Racah parameter B of 600 cm$^{-1}$ (Refs.[7,8]). The $^2T_{1g}$ ($^2A_{2g}$) state has an energy of 22B at the HS to LS transition, while the energy of the $^2E_g$ state is significantly higher (~ 29B at the spin transition). Accordingly, we obtain frequencies of 13200 and 17400 cm$^{-1}$ for the two lowest energy excited states at the spin transition pressure, which was theoretically predicted to be at ~ 30 GPa (Ref.[5]). Assuming a reasonable blue-shift of 30-90 cm$^{-1}$/GPa (Ref.[9,10]) we obtain 16200-22200 and 20400-26400 cm$^{-1}$ for the $^2T_{1g}$ ($^2A_{2g}$) and $^2E_g$ bands, respectively. Therefore, the band centered at 17500 cm$^{-1}$ can only be assigned to the $^2T_{2g} \rightarrow {}^2T_{1g} ({}^2A_{2g})$ transition in LS $Fe^{3+}$.

# SUPPLEMENTARY FIGURES

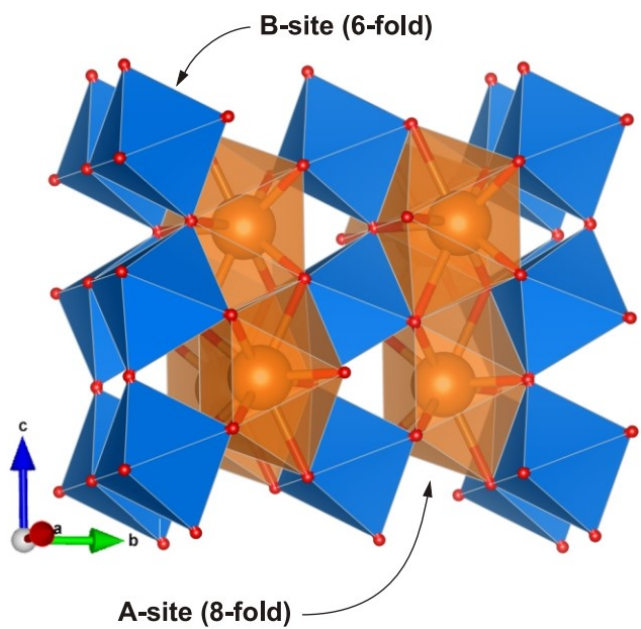

**Supplementary Figure S1.** Post-perovskite crystal structure. A-site is shown by orange polyhedrons with spheres and B-site by blue octahedra. Small red spheres represent oxygen atoms.

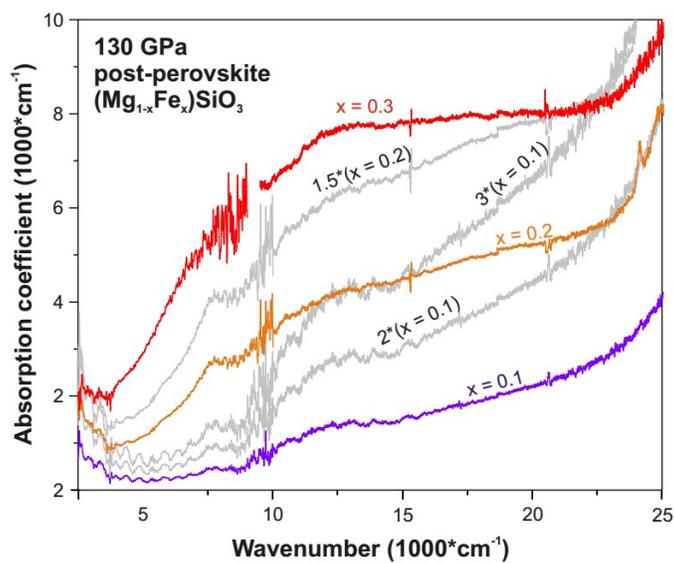

**Supplementary Figure S2.** Absorption coefficients of $Mg_{1-x}Fe_xSiO_3$ - Ppv for x=0.1 (purple), x=0.2 (orange), and x=0.3 (red), respectively (after Ref.[11]). Grey curves depict absorption coefficients of Ppv with x=0.1 and x=0.2 multiplied by 2 and 3 (for x=0.1), and by 1.5 (for x=0.2) for better comparison.

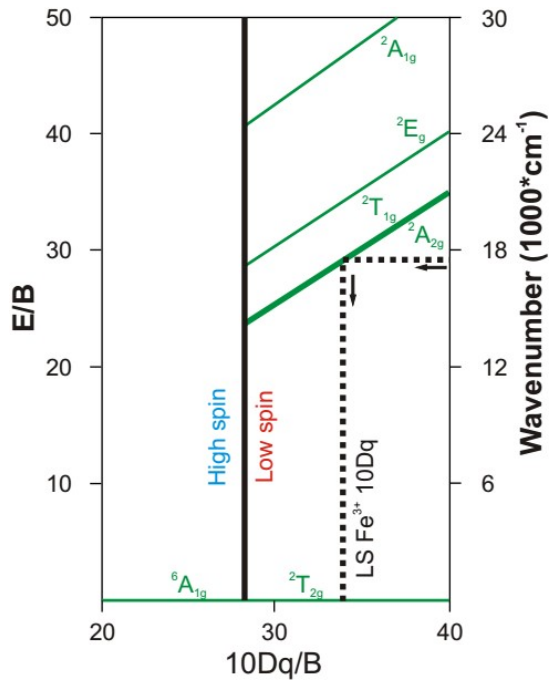

**Supplementary Figure S3.** Tanabe-Sugano diagram for $d^5$ elements in 6-fold coordination. Green lines depict spin-allowed transitions as a function of the crystal field splitting energy (10Dq). The Racah parameter B is assumed to be 600 cm$^{-1}$ as the reported values are B~590-650 cm$^{-1}$ for Fe$^{3+}$ in 6-fold coordination[8]. Note that $^2T_{1g}$ and $^2A_{2g}$ excited states have identical energies. We evaluate the crystal field splitting energy of LS Fe$^{3+}$ as 10Dq = 20350 cm$^{-1}$ (the procedure is shown by the thick dotted line).

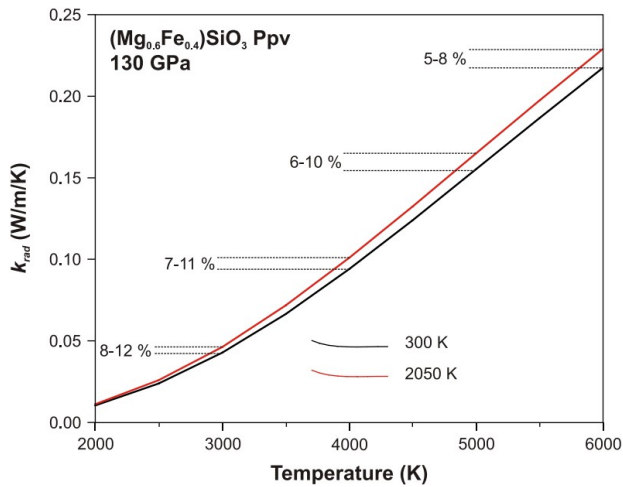

**Supplementary Figure S4.** Radiative thermal conductivity of Mg$_{0.6}$Fe$_{0.4}$SiO$_3$ - Ppv at 130 GPa evaluated using absorption coefficients shown in Fig.1 for different sample temperatures. The difference in $k_{rad}$ evaluated from the absorption spectrum at 2050 K is 5-12 % higher than that evaluated from the spectra at 300 K.

# SUPPLEMENTARY TABLES

**Supplementary Table S1.** Lattice and radiative thermal conductivities of major lower mantle minerals at expected core-mantle boundary conditions. The values used in the modelling are shown in bold.

| Phase | Mechanism | k, W/m/K | P, GPa / T, K | Composition | Source | Method |
|---|---|---|---|---|---|---|
| **Post-perovskite** | **lattice** | **16.8 ± 3.7** | **135 GPa / 3700 K** | **MgSiO$_3$** | **Ref.[12]** | **Thermoreflectance** |
| Post-perovskite | lattice | 15.1 ± 0.9 | 140 GPa / 3000 K | MgSiO$_3$ | Ref.[13] | *ab initio* |
| Post-perovskite | lattice | 10.7 | 135 GPa / 4000 K | MgSiO$_3$ | Ref.[14] | *ab initio* |
| | | | | | | |
| **Bridgmanite** | **lattice** | **9.0 ± 1.6** | **135 GPa / 3700 K** | **MgSiO$_3$** | **Ref.[12]** | **Thermoreflectance** |
| Bridgmanite | lattice | 2.3 | 100 GPa / 4000 K | MgSiO$_3$ | Ref.[15] | *ab initio* |
| Bridgmanite | lattice | 1.6 | 136 GPa / 4000 K | MgSiO$_3$ | Ref.[16] | *ab initio* |
| Bridgmanite | lattice | 6.8 ± 0.9 | 136 GPa / 4000 K | MgSiO$_3$ | Ref.[17] | *ab initio* |
| Bridgmanite | lattice | 12.4 ± 2.0 | 137 GPa / 3000 K | MgSiO$_3$ | Ref.[13] | *ab initio* |
| Bridgmanite | lattice | 7.9 | 135 GPa / 4000 K | MgSiO$_3$ | Ref.[14] | *ab initio* |
| | | | | | | |
| **Periclase, single crystal** | **lattice** | **21.5 ± 5** | **135 GPa / 3700 K** | **MgO** | **Ref.[18]** | **Thermoreflectance** |
| Periclase, powder | lattice | 17.9 ± 1.1 | 135 GPa / 3600 K | MgO | Ref.[19] | Thermoreflectance |
| Periclase | lattice | 20 ± 5 | 136 GPa / 4100 K | MgO | Ref.[20] | *ab initio* |
| | | | | | | |
| **Post-perovskite** | **radiative** | **1.2 ± 0.2** | **130 GPa / 4000 K** | **Mg$_{0.9}$Fe$_{0.1}$SiO$_3$** | **This study** | **Optical absorption at high T** |
| **Bridgmanite** | **radiative** | **2.2 ± 0.4** | **133 GPa / 4000 K** | **Mg$_{0.9}$Fe$_{0.1}$SiO$_3$** | **This study** | **Optical absorption at 300 K, corrected for high T** |
| **Ferropericlase** | **radiative** | **0.2 ± 0.1** | **133 GPa / 4000 K** | **Mg$_{0.85}$Fe$_{0.15}$O** | **This study** | **Optical absorption at 300 K** |